\documentclass[a4paper,fleqn]{cas-sc}

\usepackage[authoryear]{natbib}

\begin{document}
\let\WriteBookmarks\relax
\def\floatpagepagefraction{1}
\def\textpagefraction{.001}

\shorttitle{Optimal starting point for time series forecasting}

\shortauthors{Zhong et~al.}

\title [mode = title]{Optimal starting point for time series forecasting}



%

\author[1]{Yiming Zhong}
\ead{2023210958@email.cufe.edu.cn} 
\credit{Conceptualization of this study, Methodology, Software} 
\affiliation[1]{organization={School of Statistics and Mathematics, Central University of Finance and Economics},
	city={Beijing},
	postcode={100081},
	country={China}}

\author[2]{Yinuo Ren}
\ead{renyinuo@amss.ac.cn} 
\credit{Conceptualization of this study, Methodology, Software} 
\affiliation[2]{organization={Academy of Mathematics and Systems Science, Chinese Academy of Sciences},
	city={Beijing},
	postcode={100190},
	country={China}}

\author[3]{Guangyao Cao}
\ead{caoguangyao@cufe.edu.cn} 
\credit{Computing, Software} 
\affiliation[3]{organization={Smart Campus Construction Center, Central University of Finance and Economics},
	city={Beijing},
	postcode={100081},
	country={China}}

\author[4]{Feng Li}[orcid=0000-0002-4248-9778]
\ead{feng.li@gsm.pku.edu.cn} 
\credit{Conceptualization of this study, Methodology, Software} 
\affiliation[4]{organization={Guanghua School of Management, Peking University},
	city={Beijing},
	postcode={100871},
	country={China}}
\fnmark[1]

\author[5]{Haobo Qi}[orcid=0009-0007-8128-6360]
\ead{haobo4869@bnu.edu.cn} 
\credit{Conceptualization of this study, Methodology} 
\affiliation[5]{organization={School of Statistics, Beijing Normal University},
	city={Beijing},
	postcode={100875},
	country={China}}
\fnmark[2]
\cormark[1]
\cortext[1]{Corresponding author}

\fntext[fn1]{Feng Li is supported by the National Social Science Fund of China (No. 22BTJ028).}




\begin{abstract}
Recent advances on time series forecasting mainly focus on improving the forecasting models themselves. However, when the time series data suffer from potential structural breaks or concept drifts, the forecasting performance might be significantly reduced. In this paper, we introduce a novel approach called Optimal Starting Point Time Series Forecast (OSP-TSP) for optimal forecasting, which can be combined with existing time series forecasting models. By adjusting the sequence length via leveraging the XGBoost and LightGBM models, the proposed approach can determine the optimal starting point (OSP) of the time series and then enhance the prediction performances of the base forecasting models. To illustrate the effectiveness of the proposed approach, comprehensive empirical analysis have been conducted on the M4 dataset and other real world datasets. Empirical results indicate that predictions based on the OSP-TSP approach consistently outperform those using the complete time series dataset. Moreover, comparison results reveals that combining our approach with existing forecasting models can achieve better prediction accuracy, which also reflect the advantages of the proposed approach.
\end{abstract}




\maketitle

\section{Introduction}

Time series forecasting plays a crucial role in practical applications across various fields including services, tourism, finance, meteorology and many others. Before the rise of machine learning — particularly deep learning — traditional forecasting methods predominantly relied on statistical models like Exponential Smoothing (ETS) \citep{hyndman2008forecasting} and ARIMA \citep{box1994time}. These models, which use past observations to construct linear functions for predicting future trends, have been widely employed in forecasting tasks for decades. In recent years, deep learning techniques, such as Recurrent Neural Networks (RNNs) \citep{lipton2015critical} and ConvTrans \citep{li2019enhancing}, have gained popularity in time series forecasting. These approaches excel at capturing complex nonlinear patterns in the data, leading to more accurate predictions.

Previous literature on time series modeling has mainly concentrated on model selection and optimization, such as enhancing forecast accuracy through the integration of forecasting, ensemble learning, and artificial neural networks. These techniques rely on learning from the entire dataset to make predictions. Typically, providing a predictive model with all available data allows it to form a comprehensive understanding of historical patterns, thereby enhancing the model’s predictive capabilities. With recent advances in data storage technologies, we can now easily store and retrieve vast amounts of time series data for model training. However, in many practical applications, the distribution and trends of data are likely to change over time. There may be sudden shifts at specific time points or gradual, continuous changes over time. These phenomena are referred to as structural breaks or concept drift \citep{gama2014survey, vzliobaite2016overview} in machine learning. Therefore, if these changes in data are not taken into account, the predictive accuracy of the model might be significantly reduced.

To illustrate the idea, consider a real example in tourism. The flow of tourists between China and Japan had been steadily increasing since 1979, with a significant surge after 2010. However, the outbreak of COVID-19 in 2020 had a profound global impact, and the subsequent quarantine measures imposed by various countries caused a sharp decline in travel. As a result, the number of Chinese tourists visiting Japan in 2020 and 2021 plummeted to levels reminiscent of the pre-2010 era. Currently, tourism is in a phase of recovery, gradually returning to the levels observed before 2019. If we use total tourist numbers data for future predictions, it might overly emphasize the consistent growth observed before 2019, failing to account for the tourism industry’s recovery from the pandemic. This could lead to significant inaccuracies in the forecast. While certain time series models, such as ETS and LSTM, give more weight to recent data to minimize the noise from long-term trends, it may be more appropriate in this case to exclude data prior to the COVID-19 outbreak to improve forecast accuracy.


However, simply truncating the data might not necessarily yield to optimal predictive results. In fact, we should expect that there exists a specific point in the time series such that if we begin the forecast from that point, the forecast error can be minimized. We refer to this as the optimal starting point (OSP) of the time series. In the previous example, we subjectively chose the starting point of a new time series based on the onset of the COVID-19 pandemic. However, for other time series data, we may not have clear prior knowledge, to pinpoint the location of structural changes that would inform data truncation. Moreover, there are various ways to truncate a time series, making it difficult to identify this point subjectively. Therefore, our goal is to develop a general framework that can help us automatically determine the OSP of a time series, especially for time series with structural breaks or concept drifts. This is the primary focus of this article.

In this paper, we propose a novel approach called Optimal Starting Point Time Series Forecast (OSP-TSP) for optimal time series forecasting. The proposed OSP-TSP method contains two main steps. In the first step, the proposed method capture the intrinsic characteristics of time series data. Then the proposed method can determine an OSP via leveraging some forecasting models. In this paper, we use the XGBoost and LightGBM \citep{GoodsForecast2024} models. In the second step, existing forecasting models can be applied to achieve the final prediction results. It is worth noting that different time series are likely to vary in length, which brings challenge for model training. When feeding a set of time series into a model, the length inconsistency can complicate the training process. To address this issue, we consider extracting consistent features from each time series before putting them into the model for training. These features, such as length, frequency, and other relevant metrics, provide important information into the characteristics of the time series and play a crucial role in improving forecasting accuracy. The performances of the OSP-TSP approach are then evaluated across various frequencies on the M4 dataset and other real-world datasets. Empirical results indicate that predictions based on the OSP-TSP approach consistently outperform those using the complete dataset. Moreover, comparison results reveals that combining our approach with some existing forecasting models can achieve better prediction accuracy, which also reflect the advantages of the proposed method.


The rest of the paper is organized as follows. Section 2 introduces the related works from optimal starting point determination and time series feature extraction. Section 3 introduces the base prediction model and method for improving predictions. Section 4 demonstrates the performances of the proposed method through empirical studies on M4 data and other datasets. Finally, Section 5 conclude this article with some discussions.

\section{Related Work}

\subsection{Optimal starting point determination}
The determination of the optimal starting point of time series is similar to the change point detection. Change point detection algorithms are a suite of methods designed to identify when significant shifts occur in time series data. The existing literature of change point detection can be categorized into supervised and unsupervised approaches. The supervised approaches include decision trees \cite{zheng2008learning}, hidden Markov models, Gaussian mixture models \cite{han2012comprehensive}, and many others. These models are frequently employed in transportation domains, such as using accelerometer or audio data to detect changes in human activity states (e.g., walking versus running). Unsupervised learning approaches, on the other hand, do not require training data. Instead, they detect patterns and changes directly from the time series data. These methods are suitable for unlabeled data and can be used without prior knowledge. These methods typically detect changes by computing or estimating a particular characteristic value of time intervals. For instance, \cite{kawahara2011sequential} proposes directly estimating the ratio of probability densities and calculating the likelihood ratio between reference and test intervals to detect changes. Similarly, \cite{liu2013change} introduces a novel statistical change point detection algorithm based on relative Pearson divergence, which accurately and efficiently estimates the non-parametric divergence between time series samples using direct density ratio estimation. Additionally, a change point detection algorithm based on subspace identification has been proposed, leveraging the approximate equivalence between the column space of the observability matrix and the space spanned by sub-sequences of the time series data \cite{kawahara2008change}. This approach evaluates incoming sub-sequences of data to achieve effective change point detection.

The proposed method is also closely related to the field of concept drift as we mentioned previously. Researchers have proposed several solutions in the past literature to address concept drift in machine learning and time series forecasting. According to \citet{Liu2023}, previous efforts can be mainly categorized into two directions. The first research direction is to detect structural breaks or concept drifts, and then process them to enhance the predictive accuracy of time series models. An intuitive idea is to combine the change point detection techniques with different time series forecasting models, see for example \citep{hadad2017improving, wan2024tcdformer, oh2000using}. Another approach is to construct a concept drift detector or filter. Once the prediction metrics reach an alert level, the detector will prompt the model to retrain using the buffered contextual data; see for example \citep{widmer1996learning, gonccalves2013rcd}. The second research direction focuses on handling concept drifts in an implicit manner. These methods typically aim to establish robust forecasting models that can withstand the effects of concept drift. For instance, \citet{kolter2007dynamic} proposed an ensemble learning framework, which can handle concept drift by dynamically adjusting and weighting the base learners in the candidates pool. \citet{Liu2023} considered establishing a forecasting model based on two subsequences, which dynamically adjusts the weights of sequence samples.

\subsection{Time series feature extraction}
Traditional time series feature extraction approaches are based on statistical methods, such as summary statistics (e.g., mean, variance, skewness, and kurtosis) \cite{hamilton1994time}.These features have expanded over time to address more complex needs. Statistical methods are simple, intuitive, and easy to implement. With the growing demand for time series analysis, several software packages have emerged for automated feature extraction. For instance, \cite{christ2017} introduced the tsfresh open-source library in Python, which implements 63 feature extraction methods, including statistical, correlation, and stationarity features. It is based on extensible hypothesis testing (like the FRESH algorithm) and automatically identifies and extracts the most relevant features for tasks such as classification or regression, making it scalable and suitable for large-scale time series data. Similarly, \cite{barandas2020tsfel} introduced TSFEL, a Python package offering over 60 feature extraction methods across time, statistical, and frequency domains. TSFEL allows customization through an online interface or directly as a Python package, making it ideal for rapid exploratory data analysis. Additionally, it provides an evaluation of computational complexity, helping users estimate the computational cost of feature extraction early in the machine learning pipeline.

There are also other ways to construct time series features. \citet{moerchen2003} proposes a novel method for time series feature selection based on DWT and DFT, which outperforms traditional methods in terms of energy preservation. This method has the potential to improve clustering results and reduce the size of classification rule sets. \cite{zhang2006unsupervised} presents an unsupervised feature extraction method based on orthogonal wavelet transform. This method automatically selects the best wavelet decomposition scale to extract key features from time series, with a time complexity of $O(mn)$, where $m$ is the number of time series and $n$ is the length of each time series. Experimental results show that the features extracted by this method can improve the quality of time series clustering. \cite{olszewski2001generalized} proposes a generalized feature extraction method for structural pattern recognition in time series data. Based on six types of modulation commonly used in signal processing, this method designs six structural detectors to identify these fundamental structures in time series. In order to analyze which features will affect the determination of the optimal starting point and improve the interpretability of the model, we choose to use the traditional method of constructing features.

\section{Methodology}

The Optimal Starting Point Time Series Forecast (OSP-TSP) algorithm contains two main steps. First, we train a supervised model (OSP) to compute the optimal starting point of the time series. Second, the optimal starting point is then used to truncate the new sequence data for the final prediction model. The key idea of the OSP-TSP algorithm is to utilize machine learning techniques to identify key characteristics of a time series, thereby predicting the interval where the optimal starting point lies. These predictions are then used to improve the accuracy of time series forecasts. Essentially, this model can be applied to any time series data to enhance prediction precision. For example, it can be employed on simulated time series generated by methods like GRATIS \cite{Kang2019GRATISGT}. Next, we will provide a detailed introduction to the proposed Algorithm.

\subsection{Simplified Interval Forecast}
We define the optimal starting point, denoted as $y_{bst}$, of a time series in the following manner. Given a complete time series dataset  $\{y_1,y_2,... ,y_T\}$, our objective is to generate a prediction for the subsequent $h$ periods that minimizes the target forecast error metrics (such as mean absolute scaled error, root mean squared error, etc.). We randomly select a point, $y_{st}$, from the sequence to serve as the starting point of a new sequence. This new sequence, ${y_{st},y_{st+1},... ,y_T}$, is formed by truncating the original data. We then feed this new sequence into the time series prediction model $M$, forecast $h$ periods into the future, and compute the final target error. The optimal starting point, $y_{bst}$, is the point that yields the smallest target error among all possible $y_{st}$.
However, it’s worth noting that accurately identifying the optimal starting point of a time series is a laborious task. In the absence of any additional information, for a sequence of length $T$, we would need to construct $T-1$ forecast models and compare the resulting errors to determine the starting point that minimizes the forecast error. If our objective is to accurately predict the position of $y_{bst}$, we would need to continuously execute the aforementioned operations during the model training process to assign labels to the training set data. This precise method, however, is time-consuming and exhibits a certain degree of instability.

To mitigate the complexity of training the model while maintaining a balance between accuracy and computational cost, we have simplified the prediction goal. Instead of predicting the exact optimal starting point, we aim to predict the optimal forecast starting interval. This can be intuitively understood as the time series interval in which the optimal starting point $y_{bst}$ is located. During the model training phase, we divide each training data sequence evenly into $m$ sub-intervals. For each time series, we only calculate and store the model target prediction error starting from $n$ points between equal partitions in each sub-interval, and subsequently mark a certain interval as the location of the optimal starting point.

For each original time series, we construct $mn$ sub-time series and $mn$ sets of prediction errors for future $h$ periods. From these prediction errors, we can identify the interval $y$ where the actual optimal starting point is located, which can then be used as the label for subsequent machine learning tasks.For each time series, after calculating the target forecast error of $mn$ series, there are two methods for marking the interval where the optimal starting point of prediction is, as follows:

\begin{enumerate}[1)]
	\item For $m$ groups of intervals, we compare the $n$ sets of errors formed by $n$ points in each sub-interval, and use the minimum value of target forecast error as the error value of that sub-interval. Then we choose the smallest value as the interval y where the optimal starting point is located.
	\item For $m$ groups of intervals, calculate the average error formed by $n$ points in each sub-interval, and take the average error as the error value of that sub-interval. Then we choose the smallest value as the interval $y$ where the optimal starting point is located.
\end{enumerate}
The above two different methods of determining the interval where the actual optimal starting point is located will affect the subsequent problem of forecast accuracy, and we will analyze this effect in the empirical analysis..

\subsection{Feature Extraction}

The R package \textit{tsfeatures} provides methods to extract features from time series data, and the specific variable names and corresponding meaning descriptions are shown in Table \ref{tab1}. We use the \textit{tsfeatures} function for each time series in the M4 data to obtain the corresponding feature data   \cite{tsfeatures}.

\begin{table*}[width=.85\linewidth,cols=4,pos=h]
	\scriptsize
	\renewcommand\arraystretch{1.5}
	\centering
	\caption{Features used for time series data}
	\label{tab1}
	\begin{tabular}{cp{11cm}}
		\toprule
		\textbf{Variable} & \textbf{Description} \\
		\midrule
		frequency & Data frequency(1 for annual data, 4 for quarterly and 12 for monthly) \\
		nperiods & Set to 1 for non-seasonal data. \\
		seasonal\_periodis & The number of seasonal periods in the data (determined by the frequency of observation, not the observations themselves). \\
		trend & Strength of trend of the the STL decomposition approach. \\
		spike & The spikiness of a time series and is computed as the variance of the leave-one-out variances of the remainder component. \\
		linearity & The linearity of a time series calculated based on the coefficients of an orthogonal quadratic regression. \\
		curvature & The curvature of a time series calculated based on the coefficients of an orthogonal quadratic regression. \\
		e\_acf1 & The first autocorrelation coefficient of the the autocorrelation function of $e_t$. \\
		e\_acf10 & The sum of the first ten squared autocorrelation coefficients of the the autocorrelation function of $e_t$. \\
		seasonal\_strength & Strength of seasonality of the the STL decomposition approach. \\
		peak & The size and location of the peaks in the seasonal component. \\
		trough & The size and location of the troughs in the seasonal component. \\
		entropy & The forecastability of a time series, where low values indicate a high signal-to-noise ratio, and large values occur when a series is difficult to forecast. \\
		x\_acf1 & The first autocorrelation coefficient of the the autocorrelation function of the series. \\
		x\_acf10 & The sum of squares of the first ten autocorrelation coefficient of the the autocorrelation function of the series. \\
		diff1\_acf1 & The first autocorrelation coefficient of the  autocorrelation function of the differenced series. \\
		diff1\_acf10 & The sum of squares of the first ten autocorrelation coefficient of the the autocorrelation function of the differenced series. \\
		diff2\_acf1 & The first autocorrelation coefficient of the autocorrelation function of the twice-differenced series. \\
		diff2\_acf10 & The sum of squares of the first ten autocorrelation coefficient of the the autocorrelation function of the twiced-differenced series. \\
		seas\_acf1 & The first autocorrelation coefficient of the autocorrelation function of the seasonality. \\
		\bottomrule
	\end{tabular}
\end{table*}

\subsection{Training the Optimal Starting Point Model}
Upon extracting the features from all training data, we feed them into the Optimal Starting Point (OSP) model $C$ as independent variables. The dependent variable $y$ represents the interval in which the optimal starting point is located. The OSP model $C$ chosen for this study primarily relies on machine learning algorithms, such as XGBoost or LightGBM.
\begin{itemize}
	\item XGBoost is an efficient implementation of the Gradient Boosting Decision Tree (GBDT) and offers improvements over the original GBDT  \citep{Chen2016XGBoostAS}. As a forward addition model, its core strategy is to ensemble multiple weak learners into a strong one using boosting. The output of each decision tree is the difference between the target value and the prediction result of all preceding trees. The final result is obtained by summing up all these outputs:
	$$
	\widehat y_i^{(t)} = \sum_{k=1}^t f_k(x_i) = \widehat y_i^{(t-1)} + f_t(x_i),
	$$

	where $y_i^{(t)}$ is the prediction result for sample i after the $t$-th, $y_i^{(t-1)}$ is the prediction results for the perecious $t-1$ trees and $f_t(x_i)$ is the $t$-th DT. XGBoost successfully alleviates the overfitting problem by introducing a regularization function $\Omega (f)$:
	$$
	\Omega (f) = \gamma T + \frac{1}{2} \lambda \sum_{t=1}^{T} t^2,
	$$
	where $\gamma$ and $\lambda$ are hyper-parameters used to control the complexity, and $T$ is the number of lead nodes.

	Therefore, the objective function of XGBoost based on the GBDT optimization objective can be expressed as:
	\begin{equation*}
		\min \left[ L\{F_t(x),y\} + \Omega (f) + \zeta \right] = \min \left\{ \sum_{i=1}^N\Big(\widehat y_i^{(t)},y_i^{(t)}\Big) + \sum_{t=1}^T \Omega (f_t) + \zeta\right\}.
	\end{equation*}

	In this manner, we can iteratively optimize the loss function to achieve the sum of all Decision Tree (DT) predictions in XGBoost through the application of an additive boosting-type ensemble model.

	\item LightGBM is a framework that implements the Gradient Boosting Decision Tree (GBDT) algorithm. It supports efficient parallel training and offers advantages such as faster training speed, lower memory consumption, improved accuracy, and distributed support \citep{Ke2017LightGBMAH}. The primary motivation behind the development of LightGBM was to address the challenges faced by GBDT in handling massive data. It aims to accelerate the training of GBDT models without compromising accuracy, thereby enabling GBDT to be used more effectively and swiftly in industrial applications. Initially, LightGBM introduces a leaf-wise growth strategy with a depth limit. This strategy avoids indiscriminate treatment of leaves at the same level, which can reduce more errors and achieve better accuracy with the same number of splits, while avoiding unnecessary overhead. Furthermore, the Gradient-based One-Side Sampling (GOSS) algorithm adopts a sample reduction approach. It excludes most samples with small gradients and calculates the information gain using only the remaining samples. This approach provides a balanced algorithm in terms of reducing the amount of data and ensuring accuracy. High-dimensional data tend to be sparse, and this sparsity has inspired the design of a lossless method to reduce the dimensionality of features. As such, the Exclusive Feature Bundling (EFB) algorithm posits that the number of features can be reduced if certain features are fused and bound together.
\end{itemize}

\begin{enumerate}[1)]
	\item The dependent variable is qualitative data, with a total of $m$ categories, and the categorical prediction model is constructed.

	\item Considering that the interval where the optimal starting point is located is continuous on the time axis, the dependent variable is regarded as quantitative data and thus a regression prediction model should be constructed. Since the direct prediction results are often not integers, the closest integer on the number axis which between 1 and $m$ is taken as the interval where the optimal starting point of the prediction is located.
\end{enumerate}

Since the M4 data cover a variety of frequency data such as monthly, quarterly, and annual data, and there are tens of thousands of each frequency data, the model constructed in this paper can more adequately identify the intervals where the corresponding optimal starting points are located based on various types of time series characteristics, and make effective forecasts.

\subsection{Prediction of Optimal Starting point}
After training the Optimal Starting Point (OSP) model, we can utilize it to enhance the predictions of the basic model $M$. For a given set of data to be predicted, our improvement method is as follows:
\begin{enumerate}[1)]
	\item Extract the features of the data to be predicted and input them into the OSP model to obtain each The optimal starting interval of each time series to be predicted $y_{bst}$
	\item Based on the optimal starting interval of prediction, determine the interval position of the data to be predicted. Then, select $n$ points in the middle partition of the interval and use them as starting points to construct $n$ new sequences.
	\item Input the $n$ new sequences into the basic prediction model $M$ to obtain $n$ prediction sequences. The final prediction result is then obtained by averaging these n prediction results.
\end{enumerate}

\subsection{Evaluation Metrics}
In order to maximize prediction accuracy, the objective function of the OPS-TSP model is defined by
$$
\operatorname{argmin}\ \operatorname{Metric}(\widehat y, y, m, n),
$$
where $m$ is the number of interval of each series data, $n$ is the number of point in each sub-interval.

We select  evaluation metrics, including MASE, MAPE and calculate the average of all new time series prediction errors and compare them with the original ones.
\subsubsection{MASE}
MASE(Mean Absolute Scaled Error) is a scale-free error metric that gives each error as a ratio compared to a baseline's average error:
$$
\operatorname{MASE}=mean(|q_t|),
$$
where:
$$
q_t=\frac{e_t}{\frac{1}{n-1}\Sigma^n_{i=2}|y_i-y_{i-1}|}.
$$
\subsubsection{MAPE}
MAPE(Mean Absolute Percentage Error) is 0\% for a perfect model and greater than 100\% for an inferior model:
$$
\operatorname{MAPE}=\frac{100\%}{n}\sum_{i=1}^{n}\Big|\frac{\widehat{y_i}-y_i}{y_i}\Big|.
$$

\section{Numerical Studies}

In this section, we present several numerical studies to demonstrate the finite sample performances of the proposed OSP-TSP algorithm. Codes for reproducing the results are available at \url{https://github.com/feng-li/forecasting-with-optimal-starting-point}.

\subsection{The M4 dataset}

We have chosen the M4 dataset to validate the effectiveness of our proposed Optimal Starting Point (OSP) model construction process. The M4 dataset, compiled by the National Technical University of Athens (NTUA), comprises a total of 100,000 time series collected from several publicly accessible websites, ensuring a diverse and rich set of data sources. Furthermore, the M4 dataset spans multiple industry sectors, including services, tourism, imports and exports, demographics, education, and many others \citep{SPILIOTIS202037}. This wide coverage lends strong representation and applicability to the field of time series forecasting. The specific data types in various fields are shown in Table \ref{tab2}. Moreover, the M4 dataset is distinguished by its substantial size and comprehensive coverage, including six different data frequencies—annual, quarterly, monthly, weekly, daily, and hourly—as well as six application domains: microeconomic, macroeconomic, financial, demographic, and others. Notably, the time series in the M4 dataset are, on average, longer than those in the M3 dataset, offering more extensive data for complex methodologies for effective training \citep{Kang2017VisualisingFA}.

\begin{table}[width=.8\linewidth,cols=4,pos=h]
	\caption{Number of M4 series per data frequency and domain.\label{tab2}}
		\begin{tabular}{cccccccc}
			\toprule
			\textbf{Data Frequency} &\textbf{Micro} & \textbf{Industry} & \textbf{Macro} & \textbf{Finance} & \textbf{Demographic} & \textbf{Other} & \textbf{Total} \\
			\midrule
			Yearly & 6,538 & 3,716 & 3,903 & 6,519 & 1,088 & 1,236 & 23,000 \\
			Quarterly & 6,020 & 4,637 & 5,315 & 5,305 & 1,858 & 865 & 24,000 \\
			Monthly & 10,975 & 10,017 & 10,016 & 10,987 & 5,728 & 277 & 48,000 \\
			Weekly & 112 & 6 & 41 & 164 & 24 & 12 & 359 \\
			Daily & 1,476 & 422 & 127 & 1,559 & 10 & 633 & 4,227 \\
			Hourly & 0 & 0 & 0 & 0 & 0 & 414 & 414 \\
			Total & 25,121 & 18,798 & 19,402 & 24,534 & 8,708 & 3,437 & 100,000 \\
			\bottomrule
		\end{tabular}
\end{table}

To start with, we select three types of seasonal data (i.e. yearly, quarterly, and monthly) that have the largest volume for empirical research. For each type of data, 70\% of the data is selected as the training set to train the OSP prediction model. Specifically, the ETS and thetaf models are chosen as the basic forecasting model $M$. We consider four different improvement methods under two different class for each forecasting model. This leads to a total of sixteen improvement methods. The sequence segmentation parameters $m$ and $n$ are set to be 5 and 4, respectively. We use the MASE metric as the target error to be minimized for training the OSP model. The trained OSP model is then used to predict the optimal starting points for the remaining data. These predicted optimal starting points are used to reconstruct the new data and make predictions using ETS and thetaf models. The averaged error (i.e. MAPE and MASE) of the prediction results based on OPS-TSP method on test data are reported in Table \ref{tab3}, as well as those results that using the total time series data. Here, the MASE results, as our training target, directly reflect the prediction improvement, while the MAPE results indicate whether other non-target errors can be improved simultaneously. The best results are presented in boldface.

From Table \ref{tab3}, we find that the MASE error results obtained using our OSP-TSP method are smaller in almost all cases as compared to those prediction results obtained by using the total time series. This indicates that our method indeed yields better prediction results. Similar results can also be found in most cases in terms of MAPE results. Moreover, it would render the program overly complex if we construct eight different improvement methods to improve the prediction results. Therefore, we consider opting for only a subset of these improvement methods. The ‘mean’ column in Table \ref{tab3} represents the results of training with the minimum average error as the target. We find that training with the average minimum error as the objective consistently yields better results than training with the absolute minimum. This can be attributed to two factors. First, the instability that arises from intercepting the change of prediction error at the starting point of the time series. Second, our final prediction is constructed by taking the four equally divided points of the optimal starting interval of the prediction, and then averaging the values after prediction. This approach is also similar to the idea of minimum average error, which requires the overall prediction results to be better within an interval. Consequently, we can attempt to use only the lowest average error as a prediction.

\begin{table}[width=.9\linewidth,cols=4,pos=h]
	\caption{Prediction error results for yearly, quarterly, and monthly data with $(m,n) = (5,4)$. \label{tab3}}
		\begin{tabular}{ccccccccccc}
			\toprule
			\multirow{2}{*}{\textbf{Model}}&\multirow{2}{*}{\textbf{Class}} & \multirow{2}{*}{\textbf{Method}} & \multicolumn{2}{c}{\textbf{Yearly}} & \multicolumn{2}{c}{\textbf{Quarterly}} & \multicolumn{2}{c}{\textbf{Monthly}} & \multicolumn{2}{c}{\textbf{Mean}} \\
			\cmidrule{4-11}
			~ &~ & ~ & \textbf{MAPE} & \textbf{MASE} & \textbf{MAPE} & \textbf{MASE} & \textbf{MAPE} & \textbf{MASE} & \textbf{MAPE} & \textbf{MASE} \\
			\midrule
			\multirow{9}{*}{ETS} &\multirow{4}{*}{Actual} & xgbcls & 17.70 & 2.67 & 11.95 & 0.99 & 16.69 & 0.87 & 15.74 & 1.33 \\
			~ &~ & xgbreg & 17.08 & 2.66 & 11.95 & 0.98 & 16.66 & 0.87 & 15.57 & 1.33 \\
			~ &~ & lgbcls & 17.62 & 2.61 & 12.06 & 0.99 & 16.67 & 0.87 & 15.74 & 1.32 \\
			~ &~ & lgbreg & 16.86 & 2.68 & 11.88 & 0.98 & 16.72 & 0.87 & 15.53 & 1.34 \\
			~ &\multirow{4}{*}{Average} & xgbcls & 16.49 & \textbf{2.52} & \textbf{11.75} & 0.95 & \textbf{16.49} & 0.83 & \textbf{15.29} & \textbf{1.27} \\
			~ &~ & xgbreg & 16.65 & 2.67 & 11.83 & 0.98 & 16.72 & 0.87 & 15.47 & 1.33 \\
			~ &~ & lgbcls & \textbf{16.39} & 2.51 & 11.83 & \textbf{0.94} & 16.52 & \textbf{0.83} & 15.30 & \textbf{1.27} \\
			~ &~ & lgbreg & 16.58 & 2.71 & 11.87 & 0.98 & 16.69 & 0.87 & 15.45 & 1.34 \\
			&\multicolumn{2}{c}{Total series} & 18.06 & 3.44 & 12.20 & 1.15 & 16.83 & 0.95 & 15.96 & 1.61 \\
			\midrule
			\multirow{9}{*}{thetaf}& \multirow{4}{*}{Actual} & xgbcls & 17.20 & 2.48 & 12.20 & 1.04 & 15.76 & 0.89 & 15.21 & 1.31\\
			~ & ~ & xgbreg & 16.95 & 2.51 & 11.94 & 1.00 & 15.70 & 0.88 & 15.05 & 1.30 \\
			~ & ~ & lgbcls & 17.08 & 2.48 & 12.27 & 1.06 & 15.81 & 0.89 & 15.22 & 1.31 \\
			~ & ~ & lgbreg & 17.01 & 2.52 & 11.98 & 1.00 & 15.72 & 0.88 & 15.08 & 1.31 \\
			~ &\multirow{4}{*}{Average} & xgbcls & 17.20 & \textbf{2.42} & 11.89 & \textbf{0.98} & \textbf{15.49} & 0.85 & 14.99 & \textbf{1.26} \\
			~ &~ & xgbreg & \textbf{16.87} & 2.52 & 11.85 & 1.00 & 15.53 & 0.87 & \textbf{14.93} & 1.31 \\
			~ &~ & lgbcls & 16.98 & 2.47 & 11.92 & \textbf{0.98} & 15.50 & \textbf{0.84} & 14.95 & 1.27 \\
			~ &~ & lgbreg & 16.93 & 2.52 & 11.87 & 1.00 & 15.60 & 0.88 & 14.98 & 1.31 \\
			&\multicolumn{2}{c}{Total series} & 17.06 & 3.40 & \textbf{11.82} & 1.22 & 15.68 & 0.97 & 15.04 & 1.62 \\
			\bottomrule
		\end{tabular}
\end{table}

Next, noticing that the sub-interval division may be coarse, We consider a more refined division that dividing each sequence into an average of $m=10$ sub-intervals, while keeping $n=4$. The other settings are similar to the previous numerical studies. The prediction error results are presented in Table \ref{tab4}. One can observe that the prediction results obained by our OSP-TSP method are still superior to those obtained using the total time series. In the meanwhile, compared with the prediction results in Table \ref{tab3}, we find that the model with finer molecular intervals does not yield a significant improvement effect. This suggests that employing more sub-intervals might not necessarily lead to better results. Consequently, the use of fewer sub-intervals can also yield good prediction effects and reduce our workload in the model training phase.

\begin{table}[width=.9\linewidth,cols=4,pos=h]
	\caption{Prediction error results for yearly, quarterly, and monthly data in M4 data with $(m,n) = (10,4)$\label{tab4}}
		\begin{tabular}{ccccccccccc}
			\toprule
			\multirow{2}{*}{\textbf{Model}}&\multirow{2}{*}{\textbf{Class}} & \multirow{2}{*}{\textbf{Method}} & \multicolumn{2}{c}{\textbf{Yearly}} & \multicolumn{2}{c}{\textbf{Quarterly}} & \multicolumn{2}{c}{\textbf{Monthly}} & \multicolumn{2}{c}{\textbf{Mean}} \\
			\cmidrule{4-11}
			~ &~ & ~ & \textbf{MAPE} & \textbf{MASE} & \textbf{MAPE} & \textbf{MASE} & \textbf{MAPE} & \textbf{MASE} & \textbf{MAPE} & \textbf{MASE} \\
			\midrule
			\multirow{9}{*}{ETS} &\multirow{4}{*}{Actual} & xgbcls & 17.18 & 2.67 & \textbf{11.86} & 1.00 & 16.75 & 0.90 & 15.62 & 1.35 \\
			~ &~ & xgbreg & 17.22 & 2.70 & 11.96 & 0.98 & 16.87 & 0.91 & 15.71 & 1.36 \\
			~ &~ & lgbcls & 17.32 & 2.68 & 11.88 & 1.00 & 16.75 & 0.91 & 15.66 & 1.36 \\
			~ &~ & lgbreg & 17.00 & 2.70 & 11.93 & 0.98 & 16.89 & 0.91 & 15.66 & 1.36 \\
			~ &\multirow{4}{*}{Average} & xgbcls & 16.94 & 2.63 & 11.96 & 0.96 & 16.54 & 0.88 & \textbf{15.48} & 1.32 \\
			~ &~ & xgbreg & \textbf{16.80} & 2.67 & 11.93 & 0.97 & 16.92 & 0.91 & 15.63 & 1.35 \\
			~ &~ & lgbcls & 16.84 & \textbf{2.57} & 11.92 & \textbf{0.95} & \textbf{16.52} & \textbf{0.88} & 15.43 & \textbf{1.31} \\
			~ &~ & lgbreg & 16.88 & 2.67 & 11.99 & 0.98 & 16.92 & 0.91 & 15.66 & 1.35 \\
			&\multicolumn{2}{c}{Total series} & 18.06 & 3.44 & 12.20 & 1.15 & 16.83 & 0.95 & 15.96 & 1.61 \\
			\midrule
			\multirow{9}{*}{thetaf}& \multirow{4}{*}{Actual} & xgbcls & 17.11 & 2.63 & 12.35 & 1.07 & 15.88 & 0.92 & 15.29 & 1.37\\
			~ & ~ & xgbreg & 17.12 & 2.65 & 12.06 & 1.00 & 15.71 & 0.93 & 15.13 & 1.36 \\
			~ & ~ & lgbcls & 17.57 & \textbf{2.62} & 12.58 & 1.08 & 15.91 & 0.94 & 15.47 & 1.38 \\
			~ & ~ & lgbreg & 17.12 & 2.65 & 12.03 & 1.00 & 15.72 & 0.93 & 15.13 & 1.36 \\
			~ &\multirow{4}{*}{Average} & xgbcls & 16.84 & 2.65 & 12.00 & 1.00 & 15.95 & \textbf{0.90} & 15.16 & \textbf{1.35} \\
			~ &~ & xgbreg & 16.93 & 2.67 & 11.92 & \textbf{0.99} & \textbf{15.68} & 0.92 & 15.03 & 1.36 \\
			~ &~ & lgbcls & 16.91 & 2.65 & 12.00 & 1.00 & 15.92 & \textbf{0.90} & 15.17 & \textbf{1.35} \\
			~ &~ & lgbreg & \textbf{16.79} & 2.67 & 11.97 & 1.00 & 15.71 & 0.93 & \textbf{15.03} & 1.37 \\
			&\multicolumn{2}{c}{Total series} & 17.06 & 3.40 & \textbf{11.82} & 1.22 & 15.68 & 0.97 & 15.04 & 1.62 \\
			\bottomrule
		\end{tabular}
\end{table}

In order to demonstrate the universality of our improvement method across different frequency data, we further extend our numerical studies for weekly, daily, and hourly data in the M4 dataset. For simplicity, we only use the method of marking the interval with the minimum average error to train the prediction results after training the model. The results are presented in Table \ref{tab5}. We have observed similar prediction results, which indecates the universal effectiveness of the proposed method.

Simultaneously, we have distilled and presented the five most pivotal variables within the framework of various seasonal models, as detailed in Table \ref{tab6}. Notably, regardless of the seasonal model in question, Curvature and Linearity occupy paramount positions, forming the cornerstone features for predicting optimal starting points. For datasets exhibiting pronounced seasonality, ${\it seas\_acf1}$ (seasonal autocorrelation at lag 1) and ${\it seasonal\_strength}$ consistently rank high across quarterly, monthly, and even hourly models, emphatically demonstrating the indispensable nature of seasonal factors across these diverse timescales. Furthermore, we observe that the autocorrelation of time series, as another crucial characteristic, manifests itself in differing forms contingent upon the frequency of the data. In low-frequency annual data, the autocorrelation coefficients of the raw series, particularly ${\it x\_acf1}$ (autocorrelation at lag 1) and ${\it x\_acf10}$ (sum of squares of the first ten autocorrelation coefficients), exhibit heightened significance. However, as the data frequency escalates to weekly, daily, and even hourly levels, the autocorrelation features of different series, such as ${\it diff1\_acf1}$ (autocorrelation at lag 1 after first-order differencing) and $diff2\_acf1$ (autocorrelation at lag 1 after second-order differencing), become increasingly pivotal. In summary, for more granular time series data, the autocorrelation features of series after multiple differencing emerge as the decisive factors in identifying and determining optimal starting points.


\begin{table}[width=.7\linewidth,cols=4,pos=h]
	\caption{Prediction error results for weekly, daily, and hourly data in M4 data with $(m,n) = (5,4)$\label{tab5}}
		\begin{tabular}{cccccccc}
			\toprule
			\multirow{2}{*}{\textbf{Model}}& \multirow{2}{*}{\textbf{Method}} & \multicolumn{2}{c}{\textbf{Weekly}} & \multicolumn{2}{c}{\textbf{Daily}} & \multicolumn{2}{c}{\textbf{Hourly}} \\
			\cmidrule{3-8}
			~ &~ & \textbf{MAPE} & \textbf{MASE} & \textbf{MAPE} & \textbf{MASE} & \textbf{MAPE} & \textbf{MASE} \\
			\midrule
			\multirow{5}{*}{ETS} & xgbcls  & 9.17 & \textbf{1.98} & 4.90 & 2.71 & 22.15 & 1.79 \\
			~ & xgbreg & 9.21 & 2.00 & 5.11 & 2.77 & \textbf{21.65} & \textbf{1.61} \\
			~ & lgbcls & 9.22 & 2.00 & 4.95 & \textbf{2.69} & 21.68 & 1.72 \\
			~ & lgbreg & \textbf{9.11} & 1.99 & 5.08 & 2.77 & 22.09 & 1.71 \\
			~ & Total series & 9.17 & 2.50 & \textbf{4.81} & 3.28 & 23.15 & 1.99 \\
			\midrule
			\multirow{5}{*}{thetaf} & xgbcls & 9.19 & 2.09 & \textbf{4.97} & \textbf{2.70} & 25.04 & 1.63\\
			~ & xgbreg & 9.25 & 2.12 & 5.35 & 2.77 & 24.52 & 1.62 \\
			~ & lgbcls & 9.19 & \textbf{2.04} & 5.12 & 2.72 & 24.34 & \textbf{1.61} \\
			~ & lgbreg & 9.27 & 2.12 & 5.42 & 2.79 & 24.50 & 1.64 \\
			~ & Total series & \textbf{9.05} & 2.66 & 5.58 & 3.25 & \textbf{23.57} & 2.60 \\
			\bottomrule
		\end{tabular}
\end{table}

\begin{table}[width=.9\linewidth,cols=4,pos=h]
	\caption{The importance of the top 5 features under different seasonal data.\label{tab6}}
		\begin{tabular}{ccccccc}
			\toprule
			\textbf{Order} & \textbf{Yearly} & \textbf{Quarterly} & \textbf{Monthly} & \textbf{Weekly} & \textbf{Daily} & \textbf{Hourly}  \\
			\midrule
			1 & curvature & curvature & linearity & curvature & curvature & seas\_acf1  \\
			2 & linearity & linearity & length & linearity & linearity & curvature  \\
			3 & x\_acf1 & length & curvature & diff1\_acf1 & diff1\_acf10 & linearity  \\
			4 & trend & seas\_acf1 & seas\_acf1 & x\_acf1 & spike & diff2\_acf1  \\
			5 & x\_acf10 & seasonal\_strength & seasonal\_strength & diff1\_acf10 & length & diff2\_acf10  \\
			\bottomrule
		\end{tabular}
\end{table}


\subsection{Other real world data}

In this subsection, we consider five more real world datasets to demonstrate the finite sample performances of the proposed OSP-TSP approach. Specifically, they are GDP, Construction Industry, Exchange Rate, Confidence Index, and Import Value. As we mentioned previously, the proposed OSP-TSP approach is a general framework, which can be integrated with a variety of time series forecasting models. In this numerical experiment, we consider the ARIMA model and the neural network time series forecasting model nnetar as our base forecasting model. We still employ metrics such as Mean Absolute Scaling Error (MASE) and Mean Absolute Percentage Error (MAPE) to evaluate the prediction results of our proposed approach. Each dataset is partitioned into two parts, where 70\% of the data is allocated for training and the remaining 30\% is reserved for testing. We first use the training set to train the OSP model, and then use the trained model on the remaining dataset to predict the optimal starting point. Subsequently, we construct a new time series to improve the prediction results. The prediction results on the five real world datasets are reported in Table \ref{tab7}. One can find that the OSP-TSP approach consistently yield better prediction results than those using the total time series data under the same base models in most cases. It is worth noting that we find some models do not perform well on the exchange rate dataset. In fact, under the efficient market hypothesis, one would expect a random walk to perform best on these series, which only uses one data point. Therefore, the proposed OSP-TSP approach does not lead to consistent improvements.

\begin{table}[width=1.\linewidth,cols=4,pos=h]
	\caption{Prediction error results for five real world datasets with four different base models. The best results are in boldface.\label{tab7}}
        \resizebox{\textwidth}{!}{
		\begin{tabular}{cccccccccccc}
			\toprule
			\multirow{3}{*}{\textbf{Model}} & \multirow{3}{*}{\textbf{Method}} & \multicolumn{2}{c}{\textbf{Yearly}} &\multicolumn{2}{c}{\textbf{Quarterly}}& \multicolumn{6}{c}{\textbf{Monthly}} \\ \cmidrule(lr){3-4} \cmidrule(lr){5-6}\cmidrule(lr){7-12}
			~ & ~ & \multicolumn{2}{c}{\textbf{GDP}} & \multicolumn{2}{c}{\textbf{Industry}} & \multicolumn{2}{c}{\textbf{Exchange rate}} & \multicolumn{2}{c}{\textbf{Confidence index}} & \multicolumn{2}{c}{\textbf{Amount of imports}}   \\
			\cmidrule(lr){3-4} \cmidrule(lr){5-6} \cmidrule(lr){7-8} \cmidrule(lr){9-10} \cmidrule(lr){11-12}
			~ & ~ & \textbf{MAPE} & \textbf{MASE} & \textbf{MAPE} & \textbf{MASE} & \textbf{MAPE} & \textbf{MASE} & \textbf{MAPE} & \textbf{MASE} & \textbf{MAPE} & \textbf{MASE}  \\
			\midrule
			\multirow{5}{*}{ETS} & xgbcls & 13.89 & \textbf{3.48} & \textbf{9.00} & \textbf{1.08} & 10.71 & 1.57 & 2.08 & 1.56 & \textbf{22.39} & \textbf{1.59}  \\
			~ & xgbreg & 13.66 & 4.06 & 9.90 & 1.12 & \textbf{9.36} & \textbf{1.23} & 2.13 & 1.29 & 26.04 & 1.90  \\
			~ & lgbcls & 13.72 & 3.53 & 10.18 & 1.18 & 10.74 & 1.55 & 2.24 & 1.83 & 26.06 & 1.91  \\
			~ & lgbreg & \textbf{13.55} & 3.96 & 10.18 & 1.18 & 10.74 & 1.41 & 2.18 & \textbf{1.18} & 26.09 & 1.9  \\
			~ & Total series & 13.97 & 4.57 & 11.21 & 1.42 & 10.93 & 1.43 & \textbf{1.72} & 1.48 & 24.37 & 1.61  \\
			\midrule
			\multirow{5}{*}{thetaf} & xgbreg & 12.58 & 3.64 & 12.26 & \textbf{1.40} & 8.25 & 1.15 & 1.24 & \textbf{0.68} & 23.1 & 1.66  \\
			~ & lgbcls & 13.01 & 3.46 & \textbf{12.12} & \textbf{1.40} & 7.48 & 1.08 & 1.42 & 1.13 & 25.16 & 1.79  \\
			~ & lgbreg & 12.48 & 3.67 & \textbf{12.12} & \textbf{1.40} & 7.98 & 1.16 & 1.42 & 0.76 & 25.16 & 1.79  \\
			~ & xgbcls & \textbf{12.23} & \textbf{3.34} & 12.17 & 1.41 & 8.25 & 1.19 & \textbf{1.14} & 0.74 & 23.05 & 1.65  \\
			~ & Total series & 12.99 & 4.73 & 14.89 & 1.7 & \textbf{7.29} & \textbf{0.95} & \textbf{1.14} & 1.01 & \textbf{20.65} & \textbf{1.52}  \\
			\midrule
			\multirow{5}{*}{ARIMA} & xgbcls & 13.19 & \textbf{3.28} & 19.33 & 1.51 & 10.6 & 1.55 & 1.91 & 1.52 & 23.58 & 0.95  \\
			~ & xgbreg & 14.22 & 3.68 & 19.25 & 1.40 & 10.66 & 1.46 & 1.95 & 1.07 & 22.05 & 0.87  \\
			~ & lgbcls & 13.51 & 3.3 & \textbf{18.39} & 1.40 & 9.98 & 1.5 & 1.78 & 1.48 & 23.53 & 1.08  \\
			~ & lgbreg & 16.49 & 4.55 & 19.22 & \textbf{1.34} & 9.98 & 1.33 & 1.78 & \textbf{0.98} & \textbf{21.94} & \textbf{0.86}  \\
			~ & Total series & \textbf{12.09} & 3.93 & 18.96 & 1.73 & \textbf{9.09} & \textbf{1.2} & \textbf{1.21} & 1.08 & 29.1 & 1.97  \\
			\midrule
			\multirow{5}{*}{nnetar} & xgbcls & 14.92 & 3.84 & \textbf{22.18} & \textbf{1.82} & 9.03 & 1.44 & 1.89 & 1.18 & 36.16 & 1.41  \\
			~ & xgbreg & \textbf{14.1} & 3.99 & 27.94 & 2.47 & \textbf{8.4} & \textbf{1.29} & 1.24 & 0.65 & 30.21 & \textbf{1.24}  \\
			~ & lgbcls & 14.34 & \textbf{3.69} & 28.31 & 2.22 & 9.23 & 1.51 & 1.03 & 0.68 & \textbf{29.13} & 1.37  \\
			~ & lgbreg & 14.73 & 4.17 & 27.22 & 2.64 & 8.89 & 1.46 & \textbf{0.82} & \textbf{0.42} & 36.88 & \textbf{1.24}  \\
			~ & Total series & 15.22 & 5.11 & 26.23 & 2.59 & 9.57 & 1.31 & 0.98 & 0.84 & 42.29 & 3.03  \\
			\bottomrule
		\end{tabular}
                }
\end{table}

It is worth noting that the practitioners may not always have access to large datasets, such as the M4 dataset, for training. If we only have a small amount of data, directly extracting features for model training might not yield satisfactory results. To address this issue, we propose two approaches. The first approach involves using a pre-trained model. If we have trained the OSP model on another large dataset, we can directly use it to make predictions on the current dataset. While we advocate for the training data to closely resemble the prediction dataset, given the substantial volume of data employed in training the model in our previous experiments, it has demonstrated the ability to learn and generate corresponding predictions for analogous situations present in the features of the current dataset. The second approach is to generate a simulation dataset for training, noticing that our model has relatively flexible requirements for training datasets. For instance, the GRATIS is capable of efficiently generating new time series with controllable features  \cite{Kang2019GRATISGT}. This method can effectively augment our original dataset. We can train the model on the newly generated data and then apply it to the actual data to enhance the prediction, a strategy that has proven to be effective. We have developed two numerical experiments to demonstrate the effectiveness of the proposed two approaches. The prediction error results are displayed in Table \ref{tab8} and Table \ref{tab9}, respectively. According to \ref{tab8} and Table \ref{tab9}, we find that the MASE error results obtained using our pre-trained OSP-TSP method are smaller in almost all cases as compared to those prediction results obtained by using the total time series. This indicates that our method indeed yields better prediction results as well as our pre-trained approaches. The MAPE results did not outperform those obtained using the total time series in some cases, which is reasonable since our method does not use MAPE as the target error to be minimized.

\begin{table}[width=1.\linewidth,cols=4,pos=h]
	\caption{Prediction error results of five real world datasets via pre-trained OSP model on M4 data.\label{tab8}}
        \resizebox{\textwidth}{!}{
		\begin{tabular}{cccccccccccc}
			\toprule
			\multirow{3}{*}{\textbf{Model}} & \multirow{3}{*}{\textbf{Method}} & \multicolumn{2}{c}{\textbf{Yearly}} &\multicolumn{2}{c}{\textbf{Quarterly}}& \multicolumn{6}{c}{\textbf{Monthly}} \\ \cmidrule(lr){3-4} \cmidrule(lr){5-6}\cmidrule(lr){7-12}
			~ & ~ & \multicolumn{2}{c}{\textbf{GDP}} & \multicolumn{2}{c}{\textbf{Industry}} & \multicolumn{2}{c}{\textbf{Exchange rate}} & \multicolumn{2}{c}{\textbf{Confidence index}} & \multicolumn{2}{c}{\textbf{Amount of imports}}   \\
			\cmidrule(lr){3-4} \cmidrule(lr){5-6} \cmidrule(lr){7-8} \cmidrule(lr){9-10} \cmidrule(lr){11-12}
			~ & ~ & \textbf{MAPE} & \textbf{MASE} & \textbf{MAPE} & \textbf{MASE} & \textbf{MAPE} & \textbf{MASE} & \textbf{MAPE} & \textbf{MASE} & \textbf{MAPE} & \textbf{MASE}  \\
			\midrule
			\multirow{5}{*}{ETS} & xgbcls & \textbf{14.85} & 2.61 & 9.45 & 0.94 & 11.14 & 1.68 & 2.05 & 1.52 & 28.27 & 1.10  \\
			~ & xgbreg & 15.09 & \textbf{2.32} & \textbf{9.00} & \textbf{0.91} & \textbf{10.71} & \textbf{1.55} & 2.06 & \textbf{1.50} & 28.60 & \textbf{1.02}  \\
			~ & lgbcls & 15.41 & 2.72 & 9.22 & \textbf{0.91} & 12.47 & 1.93 & 2.11 & 1.54 & 27.50 & 1.13  \\
			~ & lgbreg & 15.11 & 2.36 & 9.16 & 0.94 & 16.31 & 2.23 & 2.10 & 1.58 & 27.58 & 1.04  \\
			~ & Total series & 16.74 & 5.13 & 11.21 & 1.42 & 11.92 & 1.74 & \textbf{2.00}& 1.57 & \textbf{27.13} & 1.60  \\
			\midrule
			\multirow{5}{*}{thetaf} & xgbcls & \textbf{12.40} & 2.23 & \textbf{11.83} & \textbf{1.14} & 9.34 & 1.46 & \textbf{1.53} & \textbf{1.07} & 28.24 & 1.13  \\
			~ & xgbreg & 12.54 & \textbf{2.06} & 11.92 & 1.16 & 9.35 & 1.26 & 1.54 & 1.17 & 28.26 & \textbf{1.08} \\
			~ & lgbcls & 13.29 & 2.36 & 12.24 & 1.16 & 9.30 & 1.46 & 1.65 & 1.19 & 30.22 & 1.19  \\
			~ & lgbreg & 12.88 & 2.13 & 11.99 & 1.17 & 9.29 & \textbf{1.25} & 1.58 & 1.20 & 28.92 & 1.12  \\
			~ & Total series  & 14.20 & 4.99 & 14.89 & 1.70 & \textbf{8.99} & 1.35 & 1.56 & 1.19 & \textbf{26.89} & 1.63  \\
			\bottomrule
		\end{tabular}
                }
\end{table}

\begin{table}[width=1.\linewidth,cols=4,pos=h]
	\caption{Prediction error results of five real world datasets via pre-trained OSP model on simulated data using GRATIS.\label{tab9}}
        \resizebox{\textwidth}{!}{
		\begin{tabular}{cccccccccccc}
			\toprule
			\multirow{3}{*}{\textbf{Model}} & \multirow{3}{*}{\textbf{Method}} & \multicolumn{2}{c}{\textbf{Yearly}} &\multicolumn{2}{c}{\textbf{Quarterly}}& \multicolumn{6}{c}{\textbf{Monthly}} \\ \cmidrule(lr){3-4} \cmidrule(lr){5-6}\cmidrule(lr){7-12}
			~ & ~ & \multicolumn{2}{c}{\textbf{GDP}} & \multicolumn{2}{c}{\textbf{Industry}} & \multicolumn{2}{c}{\textbf{Exchange rate}} & \multicolumn{2}{c}{\textbf{Confidence index}} & \multicolumn{2}{c}{\textbf{Amount of imports}}   \\
			\cmidrule(lr){3-4} \cmidrule(lr){5-6} \cmidrule(lr){7-8} \cmidrule(lr){9-10} \cmidrule(lr){11-12}
			~ & ~ & \textbf{MAPE} & \textbf{MASE} & \textbf{MAPE} & \textbf{MASE} & \textbf{MAPE} & \textbf{MASE} & \textbf{MAPE} & \textbf{MASE} & \textbf{MAPE} & \textbf{MASE}  \\
			\midrule
			\multirow{5}{*}{ETS} & xgbcls & 15.14 & 3.03 & 9.65 & \textbf{0.99} & 11.62 & 1.82 & 2.11 & 1.58 & \textbf{26.05} & 1.18  \\
			~ & xgbreg & 14.95 & 2.67 & 10.91 & 1.02 & \textbf{10.93} & \textbf{1.61} & 2.04 & \textbf{1.47} & 27.70 & 1.19  \\
			~ & lgbcls & 14.96 & 2.96 & 10.49 & 1.11 & 11.62 & 1.72 & 2.12 & 1.54 & 26.10 & 1.20  \\
			~ & lgbreg & \textbf{14.67} & \textbf{2.60} & \textbf{9.54} & 1.05 & 11.77 & 1.72 & 2.08 & 1.48 & 26.39 & \textbf{1.16}  \\
			~ & Total series & 16.74 & 5.13 & 11.21 & 1.42 & 11.92 & 1.74 & \textbf{2.00} & 1.57 & 27.13 & 1.60 \\
			\midrule
			\multirow{5}{*}{thetaf} & xgbcls & 13.96 & 2.87 & \textbf{12.03} & \textbf{1.12} & 10.09 & 1.48 & 1.72 & 1.21 & 27.62 & 1.16  \\
			~ & xgbreg & 13.17 & 2.48 & 12.25 & 1.20 & 9.08 & 1.38 & \textbf{1.54} & 1.01 & 27.56 & 1.10  \\
			~ & lgbcls & 13.45 & 2.81 & 12.15 & 1.17 & 9.28 & \textbf{1.35} & 1.61 & 1.14 & 27.86 & 1.16  \\
			~ & lgbreg & \textbf{12.77} & \textbf{2.33} & 12.25 & 1.21 & 9.22 & 1.39 & 1.58 & \textbf{0.97} & 27.53 & \textbf{1.08} \\
			~ & Total series & 14.20 & 4.99 & 14.89 & 1.70 & \textbf{8.99} & \textbf{1.35} & 1.56 & 1.19 & \textbf{26.89} & 1.63 \\
			\bottomrule
		\end{tabular}
              }
\end{table}

\subsection{Extensive experiments}

\textbf{Forecast combination.} In the empirical section, we use the interval with the smallest average error as the target for training the model, ultimately generating prediction results for four distinct settings. However, when we have no prior knowledge about the future, choosing one out of the four prediction results can be challenging. Therefore, we suggest considering the method of forecast combination. The most straightforward approach is to average all the forecast results and use them as the final forecast result. Alternatively, we can average the improved prediction results of the classification model or regression model separately and then use them as predictions. Moreover, we propose that an auxiliary meta-model can be trained to pre-assign weights to each prediction result to obtain the final combined prediction. For instance, when training the meta-model, you can input the features of the training set data, with the objective being to minimize the errors of the four prediction results of the training set data.

In Table \ref{tab10}, we report the combined prediction results using the average method for yearly, quarterly and monthly M4 data. It can be seen that although the combined forecast results are not as good as the previous optimal prediction results, they can still be effectively improved as compared to those using the total time series. This also resolves the difficulty of not knowing how to choose the final prediction. This also provides a viable approach for selecting the final prediction results.

\begin{table}[width=.7\linewidth,cols=4,pos=h]
	\caption{Prediction error results of combined prediction and random starting point.\label{tab10}}
	\begin{tabular}{ccccccccc}
			\toprule
			\multirow{2}{*}{\textbf{Model}}& \multirow{2}{*}{\textbf{Method}} & \multicolumn{2}{c}{\textbf{Yearly}} & \multicolumn{2}{c}{\textbf{Quarterly}} & \multicolumn{2}{c}{\textbf{Monthly}} \\
			\cmidrule{3-8}
			~ &~ & \textbf{MAPE} & \textbf{MASE} & \textbf{MAPE} & \textbf{MASE} & \textbf{MAPE} & \textbf{MASE} \\
			\midrule
			\multirow{5}{*}{ETS} & Full model   & \textbf{16.19} & 2.62 & \textbf{11.67} & 0.97 & 16.46 & 0.85 \\
			~ & classification   & 16.27 & \textbf{2.51} & 11.72 & \textbf{0.94} & \textbf{16.45} & \textbf{0.83} \\
			~ & Regression   & 16.53 & 2.70 & 11.82 & 0.98 & 16.66 & 0.87 \\
			~ & Random  & 17.11 & 3.04 & 11.98 & 1.10 & 16.94 & 1.00 \\
                ~ & Total series & 18.06 & 3.44 & 12.20 & 1.15 & 16.83 & 0.95 \\
			\midrule
			\multirow{5}{*}{thetaf} & Full model  & \textbf{16.82} & 2.51 & \textbf{11.79} & 0.99 & \textbf{15.44} & 0.86  \\
			~ & classification  & 16.98 & \textbf{2.41} & 11.86 & \textbf{0.97} & 15.46 & \textbf{0.84}  \\
			~ & Regression  & 16.84 & 2.53 & 11.83 & 1.01 & 15.54 & 0.88  \\
			~ & Random & 17.07 & 3.01 & 12.01 & 1.14 & 16.72 & 1.04  \\
                ~ & Total series & 17.06 & 3.40 & 11.82 & 1.22 & 15.68 & 0.97  \\
			\bottomrule
	\end{tabular}
\end{table}


\textbf{Extensive results on machine learning models.} Note that the ETS and thetaf models already give more weight to the last observations, the proposed method might not offer significant improvements when the base model performs well on the total time series. Therefore, we further consider using the machine learning models to illustrate the advantages of the proposed method, where in principle all instances are weighted equally without weighting scheme. Specifically, we consider neural network autoregressive method (nnetar) provided in the forecast package as an equally weighted approach for prediction. The experiment is conducted on M4 data. The prediction error results are displayed in Table \ref{tab11}. According to Table \ref{tab11}, We find that the prediction MASE results of the proposed OSP-TSP method outperform those derived from total time series under the same machine learning base model. And the MAPE results lead to similar conclusions. These findings highlight the effectiveness of our approach in refining prediction accuracy. Although we observe that the MAPE results on hourly data are not better than those obtained from using the total time series. It is reasonable since our optimization target error is MASE rather than MAPE.


\begin{table}[width=.9\linewidth,cols=7,pos=h]
    \caption{Prediction performance of neural network autoregressive method.\label{tab11}}
    \begin{tabular}{ccccccccccccc}
        \toprule
        \multirow{2}{*}{\textbf{Category}} & \multicolumn{2}{c}{\textbf{Yearly}} & \multicolumn{2}{c}{\textbf{Quarterly}} & \multicolumn{2}{c}{\textbf{Monthly}} \\
        \cmidrule{2-7}
        ~ & \textbf{MAPE} & \textbf{MASE} & \textbf{MAPE} & \textbf{MASE} & \textbf{MAPE} & \textbf{MASE} \\
        \midrule
        Actual mean  & \textbf{20.82} & \textbf{3.54} & 16.50 & \textbf{1.35} & 19.01 & \textbf{1.03} \\
        Average mean & 21.21 & 3.57 & \textbf{15.38} & 1.37 & \textbf{18.64} & 1.03 \\
        Total series & 21.61 & 4.06 & 15.89 & 1.57 & 19.80 & 1.15 \\
        \midrule
    \end{tabular}

    \vspace{0.3cm}

    \begin{tabular}{cccccccccc}
        \toprule
        \multirow{2}{*}{\textbf{Category}} & \multicolumn{2}{c}{\textbf{Weekly}} & \multicolumn{2}{c}{\textbf{Daily}} & \multicolumn{2}{c}{\textbf{Hourly}} \\
        \cmidrule{2-7}
        ~ & \textbf{MAPE} & \textbf{MASE} & \textbf{MAPE} & \textbf{MASE} & \textbf{MAPE} & \textbf{MASE} \\
        \midrule
        Actual mean  & \textbf{8.26} & \textbf{2.40} & \textbf{4.82} & 3.54 & 25.98 & 1.03 \\
        Average mean & 8.50 & 2.56 & 5.08 & \textbf{3.42} & 23.82 & \textbf{1.01} \\
        Total series & 9.91 & 3.86 & 5.82 & 4.08 & \textbf{14.67} & 1.08 \\
        \bottomrule
    \end{tabular}
\end{table}

\textbf{Comparison with change point detection methods.} We consider using the cpm package in R \citep{cpm_edgeR} to apply change point detection technique in our comparison experiment. It contains several change-point detection methods and we adopt the CUSUM test provided by the cpm package, which is one of the most commonly used change-point detection method \citep{healy1987note, bucher2019combining}. Specifically, we first adopt the change-point detection method based on CUSUM provided to identify potential structural breaks within the data. Once the change-point is detected, we then extract the subsequence starting from the identified change-point to the end of the series. This subsequence is then used for model training and forecasting analysis. The experiment is conducted on M4 data. The prediction error results are displayed in Table \ref{tab12}. We find that the prediction performances of the proposed OSP-TSP method outperform those derived from subsequences based on change-point detection. This may be because the purpose of change point detection methods is to seek structural changes within a sequence, rather than to achieve the best prediction results. Therefore, relying solely on change point detection methods to find the optimal subsequence sequence for forecasting might not be satisfactory.

\begin{table}[width=.8\linewidth,cols=7,pos=h]
    \caption{Prediction error results of OSP-TSP method against change point detection method on M4 data.\label{tab12}}
    \begin{tabular}{ccccccccc}
        \toprule
        \multirow{2}{*}{\textbf{Model}} & \multicolumn{2}{c}{\textbf{Yearly}} & \multicolumn{2}{c}{\textbf{Quarterly}} & \multicolumn{2}{c}{\textbf{Monthly}} \\
        \cmidrule{2-7}
        ~ & \textbf{MAPE} & \textbf{MASE} & \textbf{MAPE} & \textbf{MASE} & \textbf{MAPE} & \textbf{MASE} \\
        \midrule
        OSP best & \textbf{16.39} & \textbf{2.42} & \textbf{11.75} & \textbf{0.94} & \textbf{15.49} & \textbf{0.83} \\
        cpm-Thetaf & 16.90 & 2.97 & 12.04 & 1.18 & 15.91 & 0.97 \\
        cpm-ETS    & 17.68 & 2.98 & 12.03 & 1.12 & 17.11 & 0.95 \\
        \bottomrule
    \end{tabular}

    \vspace{0.3cm} 
    \begin{tabular}{ccccccc}
        \toprule
        \multirow{2}{*}{\textbf{Model}} & \multicolumn{2}{c}{\textbf{Weekly}} & \multicolumn{2}{c}{\textbf{Daily}} & \multicolumn{2}{c}{\textbf{Hourly}} \\
        \cmidrule{2-7}
        ~ & \textbf{MAPE} & \textbf{MASE} & \textbf{MAPE} & \textbf{MASE} & \textbf{MAPE} & \textbf{MASE} \\
        \midrule
        OSP best & \textbf{9.11} & \textbf{1.98} & \textbf{4.81} & \textbf{2.69} & \textbf{21.65} & \textbf{1.61} \\
        cpm-Thetaf & 9.19 & 2.67 & 5.62 & 3.31 & 23.81 & 2.48 \\
        cpm-ETS    & 9.34 & 9.34 & 5.28 & 3.31 & 29.44 & 1.75 \\
        \bottomrule
    \end{tabular}
\end{table}

\textbf{Combined with the FFORMA model}. As we mentioned previously, the proposed OSP-TSP approach is a general framework, which can be integrated with a variety of time series forecasting models. Then it is of great interest to study whether our OSP-TSP approach can enhance the performance of some state-of-the-arts forecasting models. For example, the FFORMA \citep{FFORMA} model, which achieved the second place in the M4 competition. We consider using the FFORMA  model as our base model to compare the results with and without the application of the our OSP-TSP approach. Specifically, we use the M4 weekly data to illustrate the finite sample performance. For the experimental setup, 30\% of the data is used to train the FFORMA model, while 40\% is employed to train the optimal prediction starting point model. The remaining 30\% of the data is used to evaluate the improvement results of the proposed method. The outputs include prediction improvements obtained by training the model with both the actual optimal starting point and the averaged optimal starting point as targets. The prediction error results are reported in Table \ref{tab13}. These results encompass predictions from sub-models representing specific starting point strategies, as well as the aggregated results for the two major categories (actual and averaged optimal starting points). The evaluation results show that nearly all predictions employing optimal starting points achieve improvements in MASE. However, the performance in MAPE is relatively less consistent, likely due to the fact that the training objective was based on MASE. This suggests that while the proposed method effectively enhances the accuracy of predictions in terms of scale-invariant measures, further refinements may be needed to optimize performance for other error metrics such as MAPE.


\begin{table}[width=.7\linewidth,cols=3,pos=h]
    \caption{Prediction error results on weekly data of M4 with base model FFORMA.\label{tab13}}
    \begin{tabular}{lccc}
        \toprule
        \textbf{Class} & \textbf{Method} & \textbf{MAPE} & \textbf{MASE} \\
        \midrule
        \multirow{4}{*}{Actual}
        ~ & xgbcls & 9.82 & 1.66 \\
        ~ & xgbreg & 10.30 & \textbf{1.64} \\
        ~ & lgbreg & 10.07 & 2.00 \\
        ~ & lgbcls & 10.48 & 1.72 \\

        \multirow{4}{*}{Average}
        ~ & xgbcls & 9.65 & 1.62 \\
        ~ & xgbreg & 10.08 & 1.61 \\
        ~ & lgbreg & 9.95 & 1.95 \\
        ~ & lgbcls & 10.01 & 1.67 \\
        \midrule
        ~ & Actual mean & 10.00 & 1.82 \\
        ~ & Average mean & 9.76 & \textbf{1.81} \\
        ~ & Total series & \textbf{9.42} & 2.00 \\
        \bottomrule
    \end{tabular}
\end{table}

\section{Conclusions and future work}

In this paper, we propose a novel OSP-TSP approach for optimal forecasting. Specifically, the proposed approach contains two main steps. In the first step, the training data is evenly divided into multiple sub-intervals. We then employ XGBoost and LightGBM models to predict the optimal starting interval, and the interval that contains the OSP is then identified. In the second step, predictions are generated from a few selected, evenly distributed points within the identified interval. These predictions are used to construct tailored sequences for the baseline model. By averaging the outputs, we achieve a final prediction result that reflects enhanced accuracy.

To demonstrate the effectiveness of the proposed method, we conducted several empirical analysis on the M4 dataset as well as several other real world datasets. The results indicate that forecasts utilizing optimal starting points yield better prediction performances as compared to those baseline models which use the total time series. We find that the proposed OPS-TSP approach can be particularly useful, when the time series data have potential structural breaks or concept drifts. This is because when time series suffer from potential structural changes or concept drift, using the complete sequence for forecasting typically fails to achieve good results. Our proposed method, however, can automatically select the subsequence that achieves the optimal forecasting performance, which also can overcome the challenges posed by structural changes or concept drifts. For instance, the stock data during the implementation of specific financial policies \citep{andreou2009structural, mahata2020identification}, and the tourism industry data before and after the COVID-19 pandemic \citep{ kourentzes2021visitor, wu2023can}.

To address the data scarcity issue, we further propose approaches that making predictions with a pre-trained OSP model, which can fully leverage the knowledge and experience of the existing method. For example, we can use an OSP model pre-trained on M4 data. Moreover, tools like GRATIS can be employed to generate simulated time series with controllable characteristics for effectively augmenting the original datasets. Models trained on these simulated datasets can then be applied to real-world data so that the prediction accuracy can be effectively enhanced.

We conclude this article with several interesting future topics. First, the proposed OSP-TSP approach involves multiple forecasting and sub-segmentation, which might be computationally demanding for larger datasets or high-frequency data. Then how to reduce the computational costs of the OSP-TSP approach becomes a problem of interest. For instance, since the calculation for subsequence are independent, a parallel computing scheme can be directly applied to reduce the computation time. Moreover, we can apply some heuristic sorting algorithms or adopt the feature based forecasting algorithms \citep{talagala2022fformpp} to reduce the computational costs. Second, this study lacks an in-depth analysis of the specific selection criteria for multiple improved results provided by the OSP approach. It is of great interest to train an additional model, which can assign weights to the different improved forecasts for a even better prediction performance. Third, this paper only conducts empirical research using a few forecasting models as the baseline models. In theory, the choice of baseline models can be more diverse and advanced. Future work should involve a more comprehensive empirical analysis along this direction, exploring a wider range of baseline models to discover additional opportunities for enhancing prediction accuracy.




\printcredits

\bibliographystyle{cas-model2-names}

\bibliography{cas-refs}



\end{document}